\documentclass[iop,apjl]{emulateapj}
\usepackage{times}
\usepackage{epsfig}
\usepackage{amsmath, amsthm, amssymb}
\usepackage[plainpages=false, colorlinks=true, anchorcolor=blue, linkcolor=blue, citecolor=blue, bookmarks=false, urlcolor=blue]
{hyperref}
\usepackage{color}
\usepackage{microtype}
\usepackage{float}
\usepackage{verbatim}
\usepackage{wrapfig}

\newcommand{\shortauth}{Swihart et al.}
\newcommand{\slugcom}{The Astronomical Journal, 153:16 (9pp), 2017 January}
\slugcomment{\slugcom}

\lefthead{\sc \footnotesize \slugcom \hfill \shortauth}
\righthead{\sc \footnotesize \slugcom \hfill \shortauth}

\usepackage{array}
\usepackage{multirow}
\usepackage{tabu}
\usepackage{graphicx}

\usepackage{import}
\usepackage{pgffor}
\usepackage{xfrac}
\usepackage{xifthen}
\usepackage{color}
\usepackage[normalem]{ulem}
\usepackage{csquotes}

\newcommand{\totstars}{1523} 
\newcommand{\fluxratio}{3}
\newcommand{\separation}{2}

\newcommand{\magmin}{$<0.0$}
\newcommand{\magmax}{$>10.0$}
\newcommand{\percentnotskiff}{$\sim 10\%$}
\newcommand{\mediandiam}{0.527}
\newcommand{\medianchi}{1.42}
\newcommand{\medianpercenterror}{6.847}
\newcommand{\medianphotpoints}{15}
\newcommand{\medianfbolpercenterr}{1.991}
\newcommand{\mediandistance}{92.68}
\newcommand{\meanreddening}{0.130}
\newcommand{\medianVmag}{5.49}
\newcommand{\starscompared}{75}
\newcommand{\avgoffset}{0.057}
\newcommand{\stddevoffset}{0.054}
\newcommand{\nstarsbeforecutoff}{2191}

\begin{document}

\title{A Catalog of Calibrator Stars for Next-Generation Optical Interferometers}

\author{Samuel J. Swihart\altaffilmark{1},
E. Victor Garcia\altaffilmark{2,3},
Keivan G. Stassun\altaffilmark{2,4},
Gerard van Belle\altaffilmark{3},
Matthew W. Mutterspaugh\altaffilmark{5},
Nicholas Elias\altaffilmark{6}}

\affil{
    \altaffilmark{1}{Department of Physics and Astronomy, Michigan State University, East Lansing, MI 48824, USA}\\
    \altaffilmark{2}{Vanderbilt University, Department of Physics \& Astronomy, 6301 Stevenson Center Ln., Nashville, TN 37235, USA}\\
    \altaffilmark{3}{Lowell Observatory, 1400 W.\ Mars Hill Rd., Flagstaff, AZ 86001, USA}\\
    \altaffilmark{4}{Fisk University, Department of Physics, 1000 17th Ave. N., Nashville, TN  37208, USA}\\
    \altaffilmark{5}{Tennessee State University, 3500 John A.\ Merritt Blvd., Nashville, TN 37209, USA}\\
    \altaffilmark{6}{OAM Solutions, LLC, 9300 Stardust Trail, Flagstaff, AZ 86004, USA}\\
}

\begin{abstract}
Benchmark stars with known angular diameters are key to calibrating interferometric observations. 
With the advent of optical interferometry, there is a need for suitably bright, well-vetted calibrator stars over a large portion of the sky. 
We present a catalog of uniformly computed angular diameters for \totstars\ stars in the northern hemisphere brighter than $V=6$ and with declinations $-15^\circ$ $<\delta<$ $82^\circ$. The median angular stellar diameter is~\mediandiam~mas.
The list has been carefully cleansed of all known binary and multiple stellar systems. 
We derive the angular diameters for each of the stars 
by fitting spectral templates to the observed spectral energy distributions (SEDs)
from literature fluxes. We compare these derived angular diameters against those measured by optical interferometry for 75 of the stars, as well as to 176 diameter estimates from previous calibrator catalogs, finding in general excellent agreement. The final catalog includes our goodness-of-fit metrics as well as an online atlas of our SED fits. The catalog presented here permits selection of the best calibrator stars for current and future visible-light interferometric observations.
\end{abstract}

\section{Introduction}
\label{sec:intro}
Optical and infrared interferometry is a powerful tool for directly measuring stellar radii, constraining binary and multiple star orbits, and observing stellar surface features such as rapid rotation or star spots in the milli-arcsecond spatial resolution regime, which is inaccessible by many other observing techniques. In the northern hemisphere there are two operational six-telescope interferometers, the Navy Precision Optical Interferometer \citep[NPOI,][]{NPOI} and the Center for High Angular Resolution Astronomy \citep[CHARA,][]{CHARA}. Each has its own assemblage of different beam combiners, such as the Visible Imaging System for Interferometric Observations \citep[VISION,][]{Ghasempour12,Garcia16} and NPOI classic \citep{NPOI}, at NPOI, and the Michigan Infrared Combiner \citep[MIRC][]{MIRCcomb}, the Precision Astronomical VIsible Observations \citep[PAVO,][]{PAVOcomb}, the Visible spEctroGraph and polArimeter \citep[VEGA,][]{VEGAcomb}, and the CLassic Interferometry with Multiple Baselines \citep[CLIMB,][]{CLIMBcomb} for the CHARA array. 

All of these beam combiners measure the basic optical/IR interferometric observables: the squared visibility and/or the closure phase and triple amplitude. To do this, these optical/IR beam combiners require that the instrument's response to a point-source during an observation is well known \citep{vanBellevanBelle05}. Typically, these calibrator stars (with known angular diameters) are interleaved with target stars throughout the observation. Without observing a calibrator star with a \textit{known} angular diameter during a sequence of observations of a target star, the squared visibility cannot be calibrated accurately. Additionally, ensuring a calibrator is not in a resolved binary helps guarantee the calibrator closure phases are zero. Thus, using calibrator stars with accurate angular diameters, which are also vetted for known binary systems, lies at the heart of accurate optical/IR interferometry. 

However, often angular diameters for calibrators are obtained in an ad-hoc basis, using a myriad of SED fitting routines or by restricting observations of target stars to ones that have nearby (typically $<15\farcs$0), bright ($V<6$) calibrator stars with interferometrically measured angular diameters. In this work, we take a \textit{uniform} approach to develop a set of carefully measured angular diameters for \totstars\ stars brighter than $V=6$ in the northern hemisphere, vetted for known binary and multiple stellar systems. 

In \S~\ref{sec:calsintro} we provide some brief background for why calibrator stars are vital to doing science using optical interferometry. 
In \S~\ref{sec:methodsdata} we describe the methods used to pick the ideal sample of calibrators and the process used to construct our final catalog. 
We utilize spectral energy distribution (SED) fitting to compute the bolometric flux of each star in our list. 
We require a low reduced $\chi^{2}$ in order to keep only stars with high quality photometry that covers the entire SED (0.2--10~$\mu$m) and which are well-fit by our spectral templates. 
Results from our SED fits are presented in \S~\ref{sec:results}. In \S~\ref{sec:disc} we compare our derived angular diameters against those measured directly by interferometers and to diameter estimates from previous calibrator catalogs. 
We provide this list of calibrator stars, together with an online atlas of the SED fits, to the greater interferometry community. 
Finally, we provide a brief summary of our calibrator sample in \S~\ref{sec:conclusion}.

\section{Importance of Calibrators with Known Angular Diameters for Interferometry}
\label{sec:calsintro}

As a source increases in apparent size, it becomes more resolved by a fixed interferometric baseline and the squared visibility amplitude decreases. Ideally, the visibility squared of a point source will always be unity. Due to instrumental visibility losses however, the visibility squared measured by a beam combiner for a point source is less than unity. The beam combiner's system visibility is: 
\begin{equation}
V_{\rm system}^{2}(t) = \frac{V^{2}_{\rm obs,cal}(t)}{V_{\rm true,cal}^{2}(t)}
\end{equation}

\noindent{where $V_{\rm obs,cal}^{2}$ is the observed visibility squared of the calibrator star, and $V_{\rm true,cal}^{2}$ is the intrinsic visibility. If a calibrator star is a point source, then $V_{\rm true,cal}^{2}=1$ and $V_{\rm system}^{2} = V_{\rm obs,cal}^{2}$. For typical observing sequences in optical and IR interferometry, observations of calibrator stars are interleaved with observations of the target star throughout the night to characterize $V^{2}_{\rm system}(t)$ over time. Therefore, the intrinsic visibility of the target star measured relative to the system visibility is:}

\begin{equation}
V_{\rm true,target}^{2}(t) = \frac{V_{\rm obs,target}^{2}(t)}{V_{\rm system}^{2}(t)}
\end{equation}

\noindent{where $V_{\rm obs,target}^{2}$ is the observed visibility of the target star. Thus, it is crucial to estimate $V_{\rm system}^{2}(t)$ at any given time. Many calibrator stars however have the potential to be slightly resolved by the longest, most scientifically interesting baselines of the interferometer, so $V_{\rm true,cal}^{2}\neq1$ for these baselines. We must compute $V_{\rm true,cal}$ for the calibrator star given its true angular diameter, $\theta_{\rm cal}$. However, $\theta_{\rm cal}$ often is not directly measured by an interferometer given that most calibrator stars are chosen because they are too small to be fully resolved by the longest baselines. Fortunately we can estimate $\theta_{\rm cal}$ from the star's reddening corrected bolometric flux and effective temperature:}

\begin{equation} 
\theta = (\rm{F}_{\rm bol} / \sigma \rm{T}_{\rm eff}^4)^{1/2}
\end{equation}

\noindent{where $\sigma$ is the Stefan-Boltzmann constant, and $\rm{F}_{bol \rm}$ has units of $\text{erg}$ $\text{cm}^{-2}$ $\text{s}^{-1}$}.

\begin{deluxetable}{llcr}
\tablecaption{Comparisons of \texttt{sedFit} angular diameter estimates for different Pickles spectral template luminosity classes.}

\tablehead{HD\# & Pickles Spectral & $\theta_{\rm SED}$ $\pm$ $\sigma$ & $\chi^2_{\rm red}$ \\
& Template & (mas) & }

\startdata
 HD174464 & F2II & 0.611 $\pm$ 0.051 & 1.40 \\ 
 & F2III & 0.667 $\pm$ 0.058 & 7.30 \\ 
 HD184010 & K0III & 0.904 $\pm$ 0.057 & 5.93 \\
 & K0IV & 0.800 $\pm$ 0.039 & 0.65 \\
 HD1671 & F5IV & 0.619 $\pm$ 0.030 & 0.81 \\
 & F5V & 0.605 $\pm$ 0.016 & 1.10 
\enddata
\label{table:lumtable}
\end{deluxetable}

\section{Methods and Data}
\label{sec:methodsdata}

\subsection{SED Fitting}
\label{sec:sedfit}

In order to obtain accurate stellar angular sizes, each star was fit with a model spectral energy distribution using the fitting routine \texttt{sedFit}, written by A.\ Boden \citep{vanBelle09,vanBelle16}. \texttt{sedFit} uses an effective temperature T$_{\rm eff}$ on input, which is mapped to a corresponding Pickles \citep{Pickles98} spectral template. \texttt{sedFit} then fits the spectral template using the photometry of a given star, to compute the bolometric flux, F$_{\rm bol}$, and reddening, A$_{\rm V}$.

\subsection{Target Selection and Catalog Construction}
\label{sec:targetselect}
We compiled a list of positions, spectral types and visual magnitudes for $\sim3000$ bright (V$_{\rm mag}<\;$6) stars in the northern hemisphere with declinations~$-15^{\circ}<\delta<82^{\circ}$~using the SIMBAD database\footnote{http://simbad.u-strasbg.fr/simbad/} \citep{Simbad}. We chose a brightness limit (V$_{\rm mag}<\;$6) given that most visible light interferometers can obtain scientifically useful data on bright stars. We also removed any stars which appear in the JMMC bad calibrator list.

\subsubsection{Removal of Binaries}
\label{sec:binary}
We purged our list of all known binary and multiple stellar systems. The visibility varies as a function of baseline with the orbital separation and flux ratio for binary systems. Hence, these systems are not suitable as calibrators. We cross-referenced our list of calibrator stars with multiple binary star catalogs to be thorough. We conservatively remove all stars in our list that have an orbit listed in the 9th Catalogue of Spectroscopic Binary orbits \citep{SB9}. We also removed all stars that were listed as binaries in the Washington Double Star Catalog (WDS, \citep{WDS}) with separations $\lesssim\;$\separation\farcs0 and with flux ratios of~$\gtrsim\;$\fluxratio~mag. We also performed a search for any interferometric binaries present in the literature that could be discarded from our list of calibrators. Finally, we removed any objects that were flagged as binaries with $\lesssim\;$\separation\farcs0 separation and with flux ratios of~$\gtrsim\;$\fluxratio~mag from Hipparcos \citep{vanLeeuwen07} or the stellar multiplicity catalog of bright stars \citep{Eggleton08}. It is possible that a few unknown binary or multiple star systems that are only detectable via interferometry (not via radial velocity) may remain in our calibrator sample. These objects will be removed from our sample as they are observed by visible-light interferometers, such as VISION \citep{Garcia16}. 

\begin{deluxetable*}{lcccccccc}
\tablecaption{A Summary of our Calibrator Catalog}

\tablehead{Total & Median $\theta_{\rm LD}$ & Median $\theta_{\rm LD}$ & Median $\rm{F}_{\rm bol}$ & Median & Median & Median \# of & Median & Median \\
Stars & (mas) & (\% Error) & (\% Error) & $V_{\rm mag}$ & $\chi_{\rm red}^2$ & Phot. Points & Dist. (pc) & $A_{V}$ }

\startdata
 \totstars & \mediandiam & \medianpercenterror & \medianfbolpercenterr & \medianVmag & \medianchi & \medianphotpoints & \mediandistance &\meanreddening  
\enddata
\label{table:summary}
\end{deluxetable*}

\begin{deluxetable*}{lccccccl}
\tablecaption{Angular Diameters for 1510 Calibrator Stars Estimated from the \emph{sedFit} Fitting Routine along with the $\chi_{\rm red}^2$ as a Quality Flag}

\tablehead{HD\# & V & f$_{\rm bol}$ $\pm$ $\sigma$ & $A_{\rm V}$ $\pm$ $\sigma$ & \# of & $\theta_{\rm SED}$ $\pm$ $\sigma$ & $\chi_{\rm red}^2$ & $\rm{SpT}$ \\
& (mag) & ($\times10^{-8}$ erg cm$^{-2}$ s$^{-1}$) & (mag) & Phot. Pts & (mas) & & }

\startdata
HD87 & 5.51 & 18.900 $\pm$ 1.041 & 0.023 $\pm$ 0.046 & 17 & 0.893 $\pm$ 0.061 & 1.42 & G5III \\
HD144 & 5.59 & 25.890 $\pm$ 0.436 & 0.205 $\pm$ 0.020 & 13 & 0.227 $\pm$ 0.041 & 6.90 & B9III \\
HD360 & 5.99 & 15.830 $\pm$ 2.135 & 0.236 $\pm$ 0.078 & 10 & 0.896 $\pm$ 0.094 & 1.08 & G9III \\
HD432 & 2.27 & 322.300 $\pm$ 6.311 & 0.085 $\pm$ 0.018 & 33 & 1.990 $\pm$ 0.115 & 1.21 & F1IV \\
HD448 & 5.53 & 22.950 $\pm$ 1.920 & 0.177 $\pm$ 0.058 & 13 & 1.110 $\pm$ 0.076 & 0.83 & K0III \\
HD560 & 5.51 & 20.050 $\pm$ 0.261 & 0.000 $\pm$ 0.020 & 12 & 0.214 $\pm$ 0.032 & 2.36 & B9V \\
HD571 & 5.03 & 34.800 $\pm$ 0.975 & 0.417 $\pm$ 0.018 & 37 & 0.656 $\pm$ 0.048 & 1.23 & F2II \\
HD587 & 5.84 & 15.420 $\pm$ 0.991 & 0.000 $\pm$ 0.053 & 18 & 0.913 $\pm$ 0.057 & 1.17 & K0III \\
HD11946 & 5.26 & 21.810 $\pm$ 1.681 & 0.000 $\pm$ 0.065 & 9 & 0.281 $\pm$ 0.021 & 1.82 & A0V \\
HD11949 & 5.69 & 18.710 $\pm$ 1.108 & 0.073 $\pm$ 0.045 & 23 & 1.010 $\pm$ 0.062 & 0.92 & K0III \\
HD11973 & 4.79 & 32.400 $\pm$ 0.962 & 0.118 $\pm$ 0.025 & 26 & 0.592 $\pm$ 0.053 & 0.99 & A9.75IV \\
HD12139 & 5.87 & 19.060 $\pm$ 2.237 & 0.421 $\pm$ 0.069 & 9 & 0.952 $\pm$ 0.065 & 0.51 & K0IV \\
HD12216 & 3.98 & 77.200 $\pm$ 2.180 & 0.066 $\pm$ 0.026 & 25 & 0.567 $\pm$ 0.045 & 1.08 & A1V \\
HD12230 & 5.38 & 22.010 $\pm$ 0.264 & 0.178 $\pm$ 0.021 & 11 & 0.494 $\pm$ 0.013 & 0.51 & F0V \\
HD12235 & 5.88 & 10.740 $\pm$ 0.074 & 0.000 $\pm$ 0.013 & 21 & 0.532 $\pm$ 0.023 & 5.87 & G1IV \\
HD12303 & 5.04 & 59.120 $\pm$ 0.681 & 0.225 $\pm$ 0.014 & 28 & 0.252 $\pm$ 0.061 & 2.73 & B7III \\
HD12339 & 5.22 & 30.030 $\pm$ 1.930 & 0.173 $\pm$ 0.045 & 19 & 1.160 $\pm$ 0.107 & 0.88 & G6.5III \\
HD12446 & 5.23 & 83.700 $\pm$ 9.946 & 1.284 $\pm$ 0.083 & 7 & 0.642 $\pm$ 0.067 & 0.95 & A2IV \\
HD12447 & 5.23 & 82.800 $\pm$ 3.100 & 0.000 $\pm$ 0.040 & 8 & 0.628 $\pm$ 0.030 & 2.20 & A2V \\
HD51000 & 5.89 & 14.640 $\pm$ 1.987 & 0.214 $\pm$ 0.092 & 9 & 0.722 $\pm$ 0.081 & 0.94 & G3III \\
HD51104 & 5.92 & 18.630 $\pm$ 0.226 & 0.000 $\pm$ 0.025 & 10 & 0.171 $\pm$ 0.029 & 1.03 & B8V \\
HD51250 & 5.00 & 54.150 $\pm$ 11.040 & 0.285 $\pm$ 0.125 & 8 & 2.030 $\pm$ 0.230 & 2.51 & K2III \\
HD51814 & 5.97 & 16.670 $\pm$ 2.874 & 0.353 $\pm$ 0.092 & 6 & 0.890 $\pm$ 0.093 & 0.96 & G8III \\
HD52312 & 5.96 & 15.170 $\pm$ 0.170 & 0.000 $\pm$ 0.027 & 10 & 0.173 $\pm$ 0.031 & 2.42 & B9III \\
HD52497 & 5.18 & 34.860 $\pm$ 4.167 & 0.317 $\pm$ 0.064 & 16 & 1.210 $\pm$ 0.104 & 1.19 & G5III \\
HD52556 & 5.74 & 26.470 $\pm$ 7.244 & 0.571 $\pm$ 0.117 & 8 & 1.200 $\pm$ 0.176 & 1.32 & K0III \\
HD225003 & 5.63 & 13.890 $\pm$ 0.079 & 0.072 $\pm$ 0.017 & 14 & 0.393 $\pm$ 0.010 & 1.35 & F0V \\
HD225180 & 5.88 & 35.530 $\pm$ 0.621 & 1.161 $\pm$ 0.018 & 12 & 0.380 $\pm$ 0.070 & 3.75 & A1.5III \\
HD225216 & 5.67 & 23.320 $\pm$ 2.593 & 0.398 $\pm$ 0.060 & 8 & 1.160 $\pm$ 0.077 & 0.98 & K1IV \\
HD225289 & 5.80 & 27.510 $\pm$ 0.402 & 0.182 $\pm$ 0.023 & 11 & 0.172 $\pm$ 0.041 & 1.66 & B7III
\enddata
\tablenotetext{}{This table is available in its entirety in machine-readable form.}
\label{table:THETABLE}
\end{deluxetable*}

\subsubsection{Spectral Types}
\label{sec:skiff}
Given that the star's assumed effective temperature is important to estimating an accurate angular diameter, we obtained spectral types for each star from the Skiff Catalogue of Stellar Spectral Classifications\footnote{http://cdsarc.u-strasbg.fr/viz-bin/Cat?B/mk} \citep{skiff}. This catalogue allowed us to determine an accurate spectral type for each star by presenting us with a history of spectral classifications for each object throughout the literature. Occasionally, some~(\percentnotskiff)~of the stars in our list did not have spectral types in the Skiff catalog. For these cases, we used the SIMBAD spectral type. For our SED fitting with \texttt{sedFit}, we used a Pickles \citep{Pickles98} spectral template for each star that was closest to the spectral type and luminosity class obtained from the Skiff catalog or SIMBAD. Pickles spectral template files did not exist for every stellar spectral type. We created these missing templates by averaging two existing, closely related source templates. For example, to create a missing template of spectral type K3V, we averaged the K2V and K4V Pickles spectral templates. We also averaged the assumed effective temperatures of the source templates to estimate the effective temperature of the missing template, with the temperature errors propagating in quadrature.

\subsubsection{Uncertainty in Spectral Types and Luminosity Classes for Catalog Stars}
Many of our bright stars are Morgan Keenan spectral standards. For these stars, we assumed no uncertainty in the spectral type. Many stars in our list had spectral types and luminosity classes from the Skiff catalog that are inconsistent across the literature. For each of these stars, we used \texttt{sedFit} with a grid of spectral types and luminosity classes. The size of the grid corresponded to the range of spectral classifications presented across the literature. The resulting fluxes, as well as the angular diameters, presented in Table~\ref{table:THETABLE}~are the values from the SED fit that yielded a reduced $\chi^2$ value closest to 1.0.

As an additional check, we tested \texttt{sedFit} using Pickles spectral templates with luminosity class I or II, II or III, III or IV, and IV or V. We note that the spectral template used made a significant difference in the estimates of the angular diameters of the giant, bright giant, and supergiant stars (luminosity classes III, II, I, Table~\ref{table:lumtable}). The angular diameter estimates for the subgiants and main sequence stars (luminosity classes IV, V) did not change significantly (Table~\ref{table:lumtable}) upon using spectral templates with different luminosity class. The spectral types listed in Table~\ref{table:THETABLE}~represent the model templates that resulted in the highest quality fit.

\begin{figure*}[t]
    \centering
    \includegraphics[width=\textwidth,trim={0 7cm 0 4cm},clip]{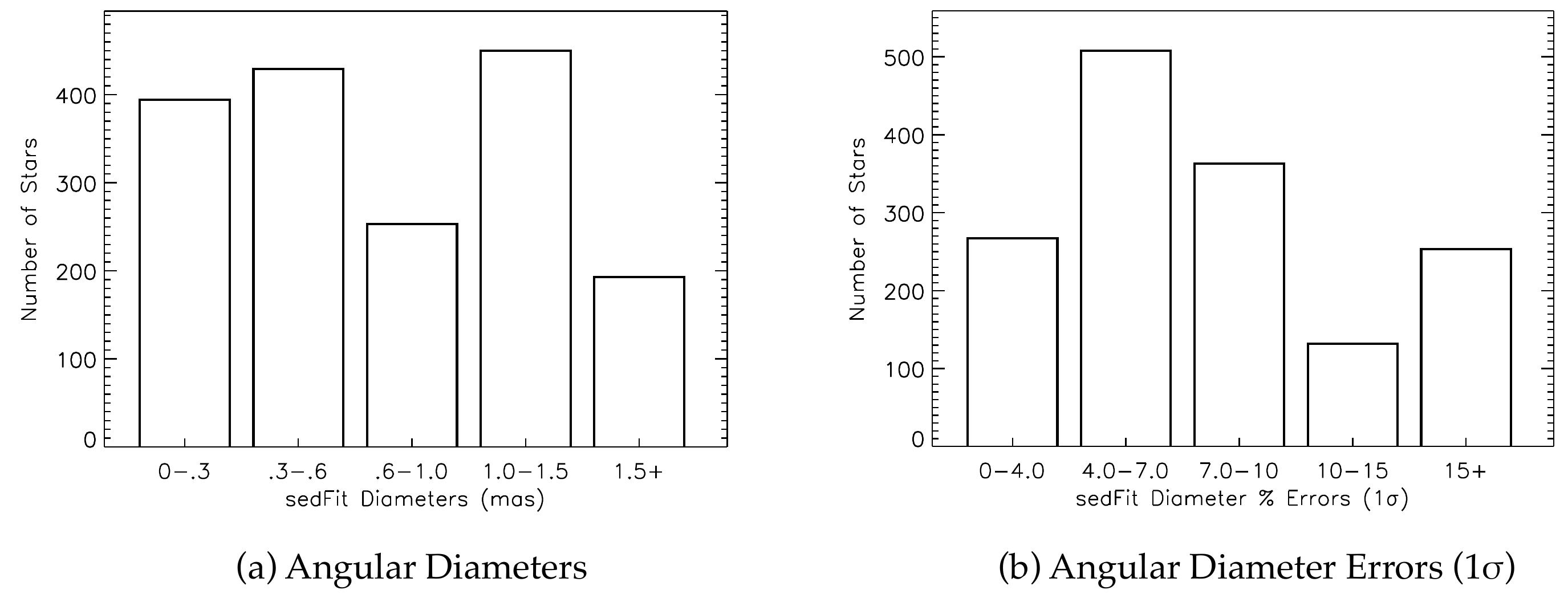}
    \caption{Distributions of angular diameters and $\rm{1}\sigma$ errors obtained using \textit{sedFit}, within our sample of~\totstars~stars. All of our calibrator stars have fitted angular diameters $<2.5$ mas, given that larger ($>2.5$ mas) stars are usually resolved by visible-light interferometers. The median error on our fitted angular diameters is $6.789$\%.}
    \label{fig:Diamdistributions}
\end{figure*}

\begin{figure*}[!htb]
    \centering
    \includegraphics[width=\textwidth,trim={0 7cm 0 4cm},clip]{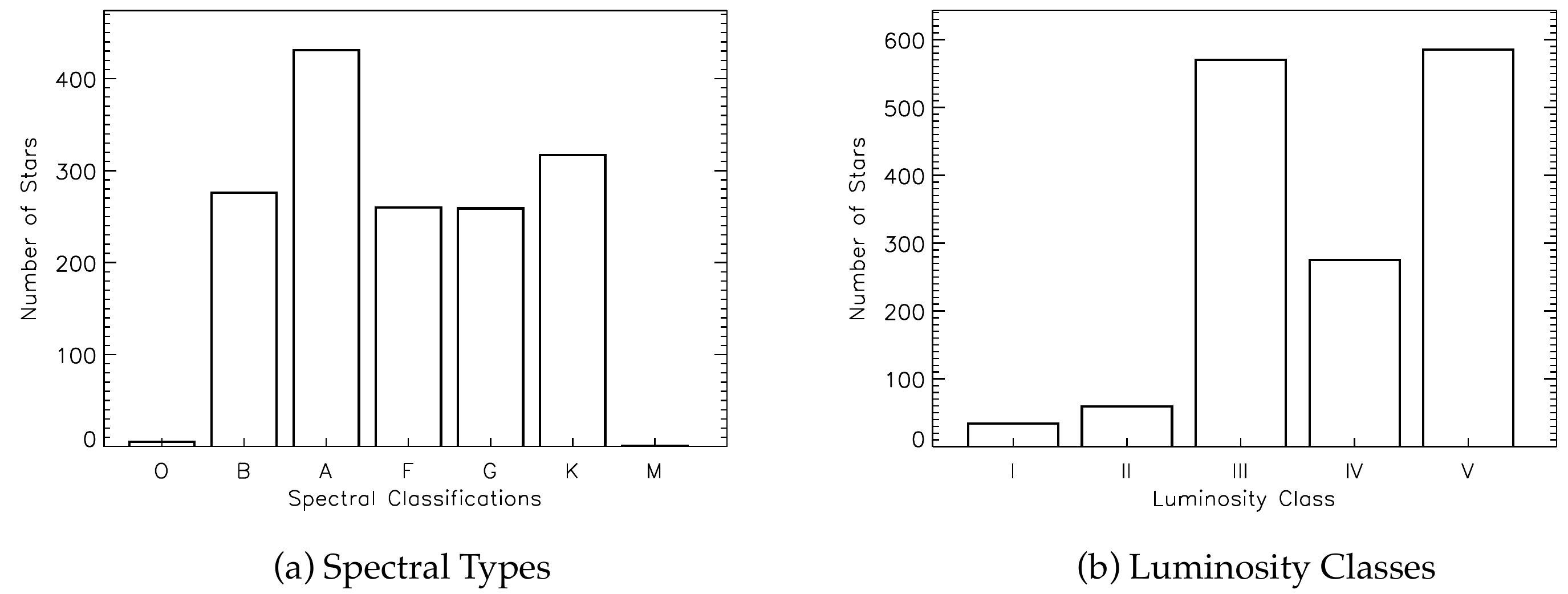}
    \caption{Distributions of spectral types and luminosity classes for our calibrator stars (see~\S\ref{sec:skiff})}
    \label{fig:distributions}
\end{figure*}

\begin{figure}[!htb]
    \centering
    \includegraphics[angle=180,width=0.54\textwidth,trim={0 0.7cm 1cm 1.5cm},clip]{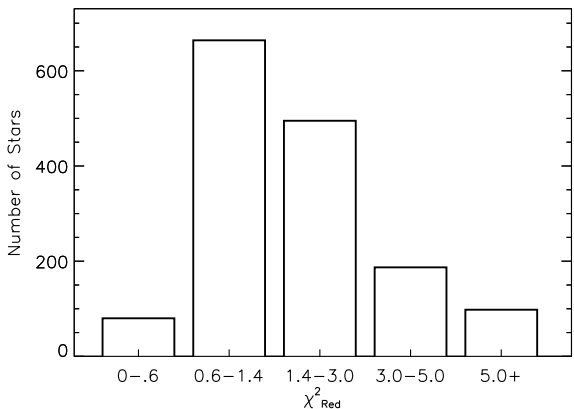}
    \caption{Reduced chi square $\chi^{2}_{\rm red}$ for the SED fits to literature photometry for each calibrator star in our sample.}
    \label{fig:chi2pdfdistribution}
\end{figure}

\subsection{Literature Photometry Quality Control}
\label{sec:photometry}

Both broadband and narrowband photometry was obtained for each star by querying both the SIMBAD \citep{Simbad} and General Catalog of Photometric Data (GCPD, \citep{Mermilliod1997}) online databases. The photometry for a given star can span $0.2-10$~$\mu$m, and contain multiple photometric systems, such as Johnson \citep{JohnsonSys}, Str\"{o}mgren \citep{StromSys}, Villinius \citep{ViliniusSys}, DDO \citep{DDOsys}, Geneva \citep{GenevaSys}, 13 color \citep{ThirteenSys}, and 2MASS \citep{2MASSsys} and IRAS \citep{IrasSys} for near to far infrared bands. 

For a small percentage of stars, photometric bandpasses had a significant amount of redundancy, due to multiple photometric catalogs including the same bandpass. These duplicated photometry measurements caused \texttt{sedFit} to improperly weight bandpasses during the fitting procedure. We removed these duplicate photometric measurements for all stars in our list. 

Furthermore, some stars had poor quality photometric measurements that were far outliers to the SED of the star. These outliers often biased the fitting procedure, yielding inaccurate angular diameters and poor quality (i.e. high reduced $\chi^{2}$ values) fits to the SED. We removed these poor quality photometry using two criteria. First we removed outliers with photometric magnitudes that were~\magmin~or~\magmax. Second, we removed photometric measurements obtained from different catalogs in the same bandpass that varied by more than $1.2\sigma$. 

Although this method of cleaning the photometry yielded much better fits to most of the stellar SEDs, there were still many fits with large reduced $\chi^{2}$ values due to problematic photometry. Therefore, we developed an objective criterion based on reduced $\chi^2$ ($\chi_\nu^2$) as follows. 
Using the set of 75 interferometric comparison stars from prior literature (see Section~\ref{sec:disc}), we found that the residuals between our estimated angular diameters and the interferometric values roughly follows the relation 
$\Delta\theta / \theta_{\rm lit} \sim 0.04 + 0.009 \times \chi_\nu^2$, where $\theta_{\rm lit}$ is the interferometrically measured angular diameter from the literature. Therefore, we opted to define an empirical threshold such that we retained only stars with $\chi_\nu^2 \le 7.0$. This corresponds to an expected fractional error on $\theta$ of at most $\sim$10\%, and represents only a mild extrapolation from the range of $\chi_\nu^2$ spanned by the 75 comparison stars (the worst outlier among the comparison set has $\chi_\nu^2 = 5.0$). 

\begin{figure*}[t]
    \centering
    \includegraphics[width=\textwidth,trim={0 7cm 0 4cm},clip]{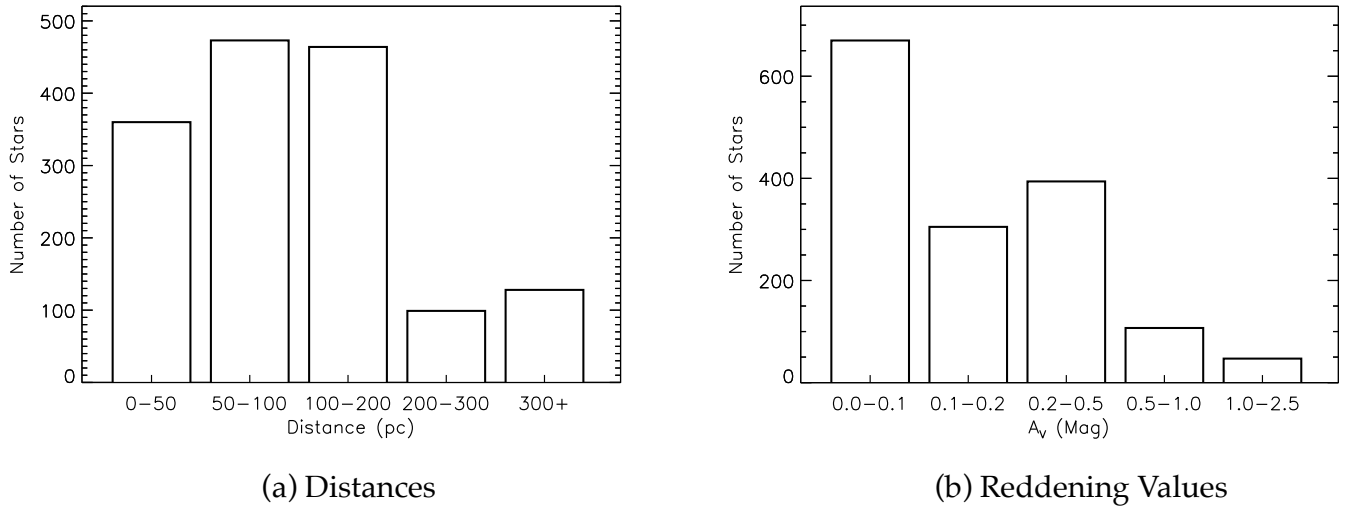}
    \caption{Distributions of Hipparcos distances and reddening values from our sample of calibrator stars. The majority of our bright $V <6$ mag stars are nearby ($<300$ pc). Therefore, most calibrator stars are expected to have low reddening $A_{\rm v}<1$.}
    \label{fig:chi2&AVdistributions}
\end{figure*}

\begin{figure*}[!htb]
    \centering
    \includegraphics[width=\textwidth,trim={0 6.7cm 0 4cm},clip]{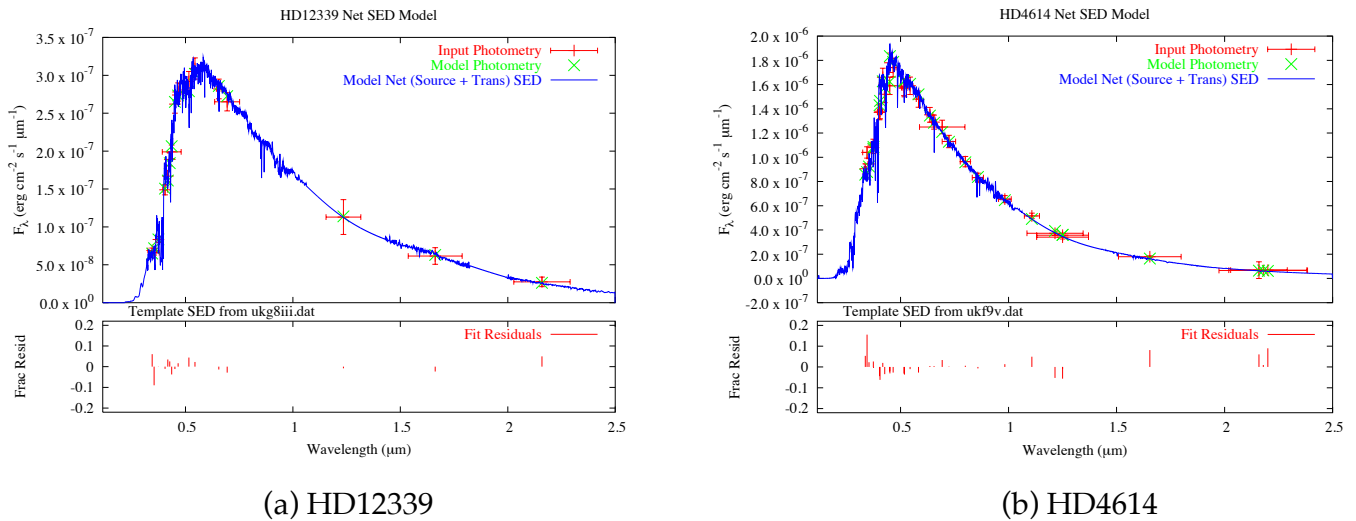}
    \caption{Example spectral energy distribution (SED) plots for HD 12339 and HD 4614. Literature photometry (red crosses) are fit by model photometry (green X's) derived from \texttt{sedFit} using a scaled, reddened spectral template (blue line) from \cite{Pickles98} (see \S\ref{sec:sedfit}) corresponding to the star's known spectral type (see \S\ref{sec:skiff}). \texttt{sedFit} minimizes the residuals between the literature photometry and the model photometry. The complete figure set (1510 images) is available in the online journal.}
    \label{fig:exampleSEDs}
\end{figure*}

\section{Results}
\label{sec:results}

We present a list of~\totstars~interferometric calibrator stars culled from all stars $V<6$ in the northern hemisphere (see \S\ref{sec:targetselect}). Angular diameters were estimated from the SED fitting routine \texttt{sedFit} (see \S \ref{sec:sedfit}). The median angular diameter $\theta_{\rm SED}$ of stars in our calibrator list is~\mediandiam~mas, while the median reduced $\chi^2_{\rm median, red}$ value for all our fits is~\medianchi~. We present the source bolometric flux, F$_{\rm bol}$, used for all fits as well as reddening, A$_{\rm v}$. A brief summary of our catalog can be found in Table~\ref{table:summary}~. The full table is presented in Table~\ref{table:THETABLE}~. The distributions of angular diameters and their errors, spectral types and luminosity classes, best-fit metrics, distances, and reddening values within our sample are displayed in Figure 1 through Figure 4 respectively. Two example SED fits for G8III giant HD12339 and F9V main sequence star HD4614 are shown in Figure~\ref{fig:exampleSEDs}~. See Appendix for fits to the SEDs for all stars in our catalog. 

\subsection{Assessing Quality of the Fits} 

From our initial large number (\nstarsbeforecutoff) of bright stars vetted for binaries for which we derived angular diameters, many had poor fits due to poor-quality photometry even after we removed much of the photometric outliers from each well-characterized SED (see \S\ref{sec:photometry}). To further improve the reliability of our calibrator star angular diameters, we first remove any stars from our calibrator catalog (Table~\ref{table:THETABLE}) that had either of the following criteria: 
\begin{itemize} 
\item Stars with $\theta_{\rm SED} > 2.5$ mas are excluded because they are resolved by most visible-light interferometers. 
\item Stars with $<4$ photometry data points are excluded because the SEDs were poorly constrained. 
\end{itemize}

These cuts removed 307 stars from our sample. Next, 
as described above, we removed any stars with reduced $\chi^{2}$ value greater than 7.0. All of the remaining stars have low reduced $\chi^{2}$ values and have been assessed to be well-fit (see Appendix for the SED fits). These checks removed an additional 361 stars, to arrive at our sample of~\totstars~total stars.

\newcommand{\nstarsstelib}{10}
\newcommand{\nstarsred}{21}
\newcommand{\medred}{0.13}
\newcommand{\massaoff}{$-0.13\pm0.17$}

\begin{figure*}[t]
    \centering
    \includegraphics[width=\textwidth,trim={0 6.3cm 0 4cm},clip]{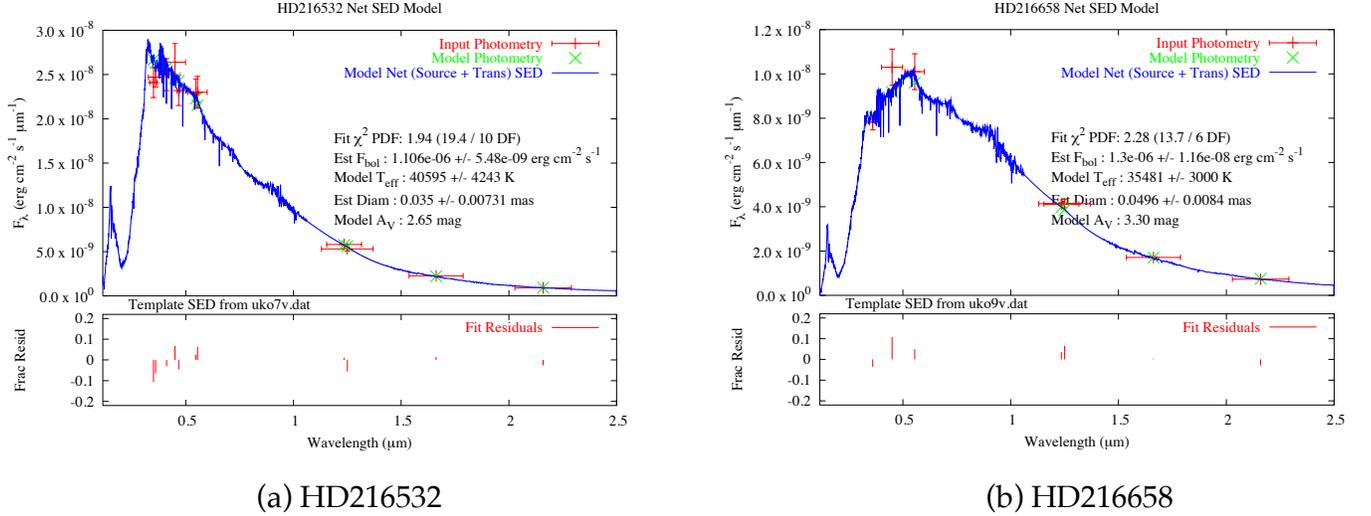}
    \caption{Example SED fits to two OB stars in the CEP-OB3 association with directly measured extinction curves by \cite{massa84}. Our best-fit reddening of A$_{\rm v}=2.65$~mag for O7V star HD216532 and A$_{\rm v}=3.30$~mag for O9V star HD216658 using \texttt{sedFit} are in good agreement with \cite{massa84} directly-observed reddening of A$_{\rm v}=2.66$ and A$_{\rm v}=3.03$, respectively.}
    \label{fig:massased}
\end{figure*}

\begin{deluxetable*}{lccccclc}
\tablecaption{We Validate our SED-Fitting Methods by Deriving the Reddening Av using \texttt{sedFit} for 21 OB Stars in the Cep OB3 Cloud.}

\tablehead{HD/ALS\# & f$_{\rm bol}$ $\pm$ $\sigma$ & $A_{\rm V}$ $\pm$ $\sigma$ & \# of & $\theta_{\rm SED}$ $\pm$ $\sigma$ & $\chi_{\rm red}^2$ & SpT & $A_{\rm v}$ \\
& ($\times10^{-8}$ erg cm$^{-2}$ s$^{-1}$) & (mag) & Phot. Pts & (mas) & & & Massa84 }

\startdata
ALS12680 & $54.33\pm0.89$ & $2.60\pm0.05$ & $10$ & $0.05\pm0.01$ & $0.85$ & B0V & $2.418$ \\
ALS12682 & $23.31\pm0.30$ & $2.75\pm0.05$ & $11$ & $0.06\pm0.02$ & $3.34$ & B2V & $3.069$ \\
ALS12732 & $18.51\pm0.27$ & $2.15\pm0.05$ & $10$ & $0.05\pm0.02$ & $1.76$ & B2V & $2.387$ \\
ALS12736 & $19.98\pm2.94$ & $2.90\pm0.15$ & $6$ & $0.06\pm0.01$ & $2.57$ & B$12$III & $2.945$ \\
ALS12766 & $18.38\pm0.45$ & $2.10\pm0.08$ & $11$ & $0.05\pm0.02$ & $4.10$ & B2V & $2.294$ \\
ALS12865 & $24.34\pm0.31$ & $1.60\pm0.05$ & $6$ & $0.07\pm0.01$ & $3.98$ & B$12$III & $1.798$ \\
ALS12867 & $76.35\pm1.05$ & $2.80\pm0.07$ & $6$ & $0.12\pm0.02$ & $0.73$ & B$12$III & $3.131$ \\
HD216532 & $110.60\pm1.35$ & $2.65\pm0.05$ & $11$ & $0.04\pm0.01$ & $1.94$ & O7V & $2.666$ \\
HD216658 & $130.00\pm2.45$ & $3.30\pm0.07$ & $7$ & $0.05\pm0.01$ & $2.28$ & O9V & $3.038$ \\
HD216711 & $35.68\pm0.37$ & $2.60\pm0.05$ & $9$ & $0.08\pm0.02$ & $3.42$ & B2V & $2.728$ \\
HD216898 & $104.00\pm20.60$ & $2.60\pm0.05$ & $12$ & $0.03\pm0.01$ & $2.92$ & O7V & $2.635$ \\
HD217061 & $52.82\pm0.44$ & $2.70\pm0.05$ & $11$ & $0.09\pm0.03$ & $5.31$ & B2V & $2.976$ \\
HD217086 & $202.80\pm42.85$ & $2.95\pm0.10$ & $8$ & $0.05\pm0.01$ & $1.13$ & O7V & $2.976$ \\
HD217297 & $64.87\pm1.12$ & $1.55\pm0.07$ & $11$ & $0.10\pm0.03$ & $3.06$ & B2V & $1.767$ \\
HD217312 & $164.40\pm1.65$ & $2.10\pm0.05$ & $7$ & $0.06\pm0.01$ & $3.71$ & O9V & $2.046$ \\
HD217463 & $28.31\pm0.48$ & $2.30\pm0.05$ & $11$ & $0.07\pm0.02$ & $1.40$ & B2V & $2.449$ \\
HD217657 & $112.50\pm2.95$ & $2.35\pm0.10$ & $9$ & $0.05\pm0.01$ & $1.69$ & O9V & $2.387$ \\
HD217979 & $23.25\pm0.27$ & $1.65\pm0.05$ & $11$ & $0.06\pm0.02$ & $4.01$ & B2V & $1.891$ \\
HD218066 & $75.37\pm2.31$ & $1.95\pm0.10$ & $7$ & $0.11\pm0.03$ & $1.12$ & B2V & $2.015$ \\
HD218323 & $96.76\pm2.59$ & $2.35\pm0.08$ & $11$ & $0.14\pm0.02$ & $1.70$ & B$12$III & $2.790$ \\
HD218342 & $87.85\pm1.72$ & $1.95\pm0.05$ & $14$ & $0.13\pm0.02$ & $2.06$ & B$12$III & $2.232$ 
\enddata
\tablenotetext{}{\textbf{Note:}Our derived A$_{\rm v}$ for each star (column 3) agree well with the directly observed A$_{\rm v}$ (last column) for these stars using the IUE satellite \citep[See Table 2, 4th column,][]{massa84}.}
\label{table:massa}
\end{deluxetable*}

\begin{figure*}[!htb]
  \centering
    \includegraphics[width=0.8\textwidth,trim={0 3cm 0 3cm},clip]{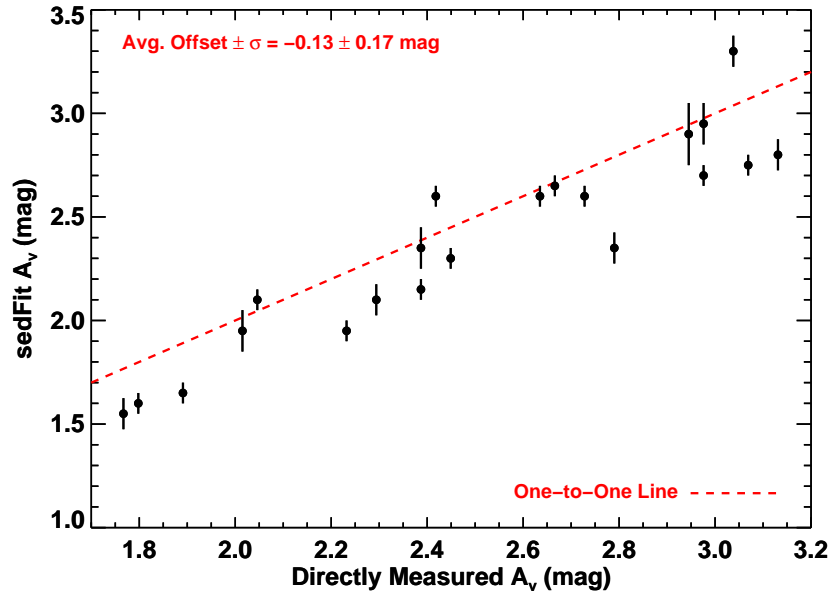}  
  \caption{We validate our SED-fitting method by comparing the reddening A$_{\rm v}$ using \texttt{sedFit} for~\nstarsred~OB stars in the Cep OB3 cloud (Y-axis) to the directly observed A$_{\rm v}$ for these stars using the IUE satellite \citep[See Table 2, 4th column,][]{massa84}. We find a statistically insignificant average offset of $-0.13\pm0.17$ mag.
}
  \label{fig:massa}
\end{figure*}

\begin{figure*}[!htb]
  \centering
    \includegraphics[width=0.8\textwidth,trim={0 3cm 0 3cm},clip]{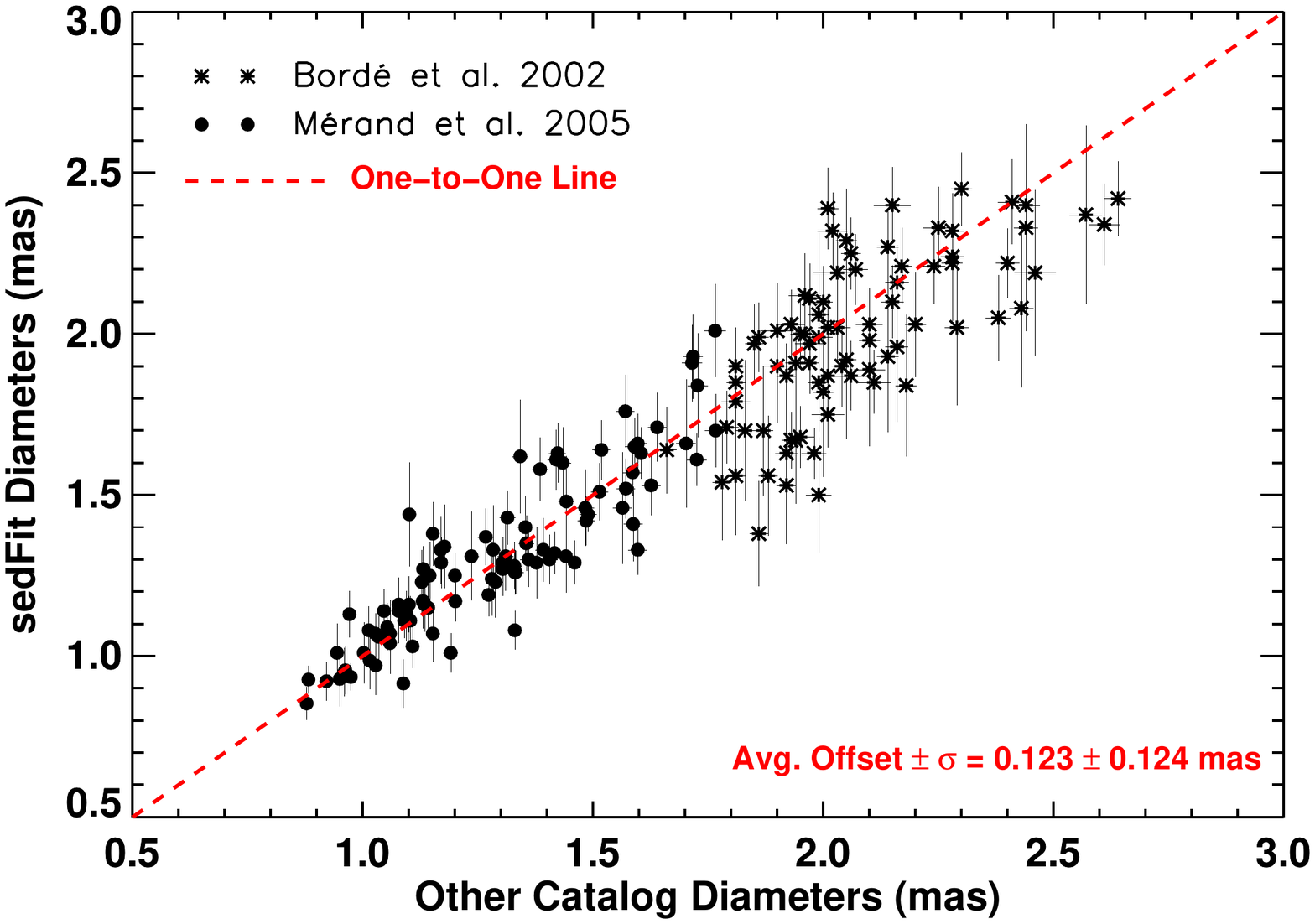}  
  \caption{Comparison of estimated angular diameters from this work and those estimated by previous calibrator catalogs from \citet{Borde:2002} and \citet{Merand:2005} (see \S~\ref{sec:disc}). There is a statistically insignificant offset of $0.123\pm0.124$ mas between our angular diameters and previous catalogs. Despite the hundreds of stars in each catalog, there are few stars in common, given that \citet{Borde:2002} and \citet{Merand:2005} optimized their catalogs for infrared interferometry -- our calibrator catalog is optimized for visible-light interferometry.
}
  \label{fig:disc}
\end{figure*}

\begin{figure*}[!htb]
  \centering
    \includegraphics[width=0.8\textwidth,trim={0 3cm 0 3cm},clip]{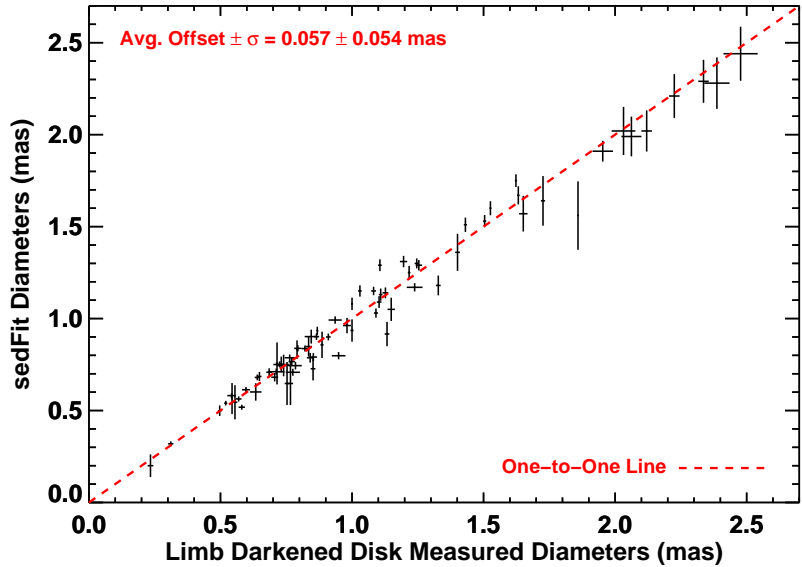}  
  \caption{Comparison of estimated angular diameters from \texttt{sedFit} and those measured interferometrically from the literature (see Table~\ref{table:compare}). Each cross is the angular diameter for a calibrator star estimated with {\it sedFit} and its corresponding directly measured limb darkened disk angular diameter from the literature.
}
  \label{fig:SEDvsMEAS}
\end{figure*}

\begin{deluxetable}{lccc}
\tablecaption{\texttt{sedFit} Angular Diameters Compared to Angular Diameters Measured by Interferometers.}

\tablehead{HD\# & $\theta_{\rm LD}$ $\pm$ $\sigma$ & $\theta_{\rm SED}$ $\pm$ $\sigma$ & REF \\
& (mas) & (mas) & }

\startdata
HD3360 & 0.311 $\pm$ 0.010 & 0.319 $\pm$ 0.013 & 4 \\
HD4614 & 1.623 $\pm$ 0.004 & 1.750 $\pm$ 0.035 & 1 \\
HD4628 & 0.868 $\pm$ 0.004 & 0.934 $\pm$ 0.022 & 2 \\
HD5015 & 0.865 $\pm$ 0.010 & 0.902 $\pm$ 0.018 & 1 \\
HD6210 & 0.520 $\pm$ 0.006 & 0.540 $\pm$ 0.013 & 3 \\
HD10476 & 1.000 $\pm$ 0.004 & 1.080 $\pm$ 0.034 & 3 \\
HD10780 & 0.763 $\pm$ 0.019 & 0.787 $\pm$ 0.018 & 1 \\
HD16160 & 1.030 $\pm$ 0.007 & 1.150 $\pm$ 0.031 & 2 \\
HD33564 & 0.640 $\pm$ 0.010 & 0.680 $\pm$ 0.016 & 8 \\
HD34411 & 0.981 $\pm$ 0.015 & 0.962 $\pm$ 0.042 & 1 \\
HD35468 & 0.715 $\pm$ 0.005 & 0.756 $\pm$ 0.114 & 6 \\
HD161797 & 1.953 $\pm$ 0.039 & 1.910 $\pm$ 0.057 & 7 \\
HD162003 & 0.949 $\pm$ 0.026 & 0.798 $\pm$ 0.020 & 1 \\
HD164259 & 0.775 $\pm$ 0.027 & 0.708 $\pm$ 0.019 & 1 \\
HD170693 & 2.120 $\pm$ 0.020 & 2.020 $\pm$ 0.112 & 5 \\
HD173667 & 1.000 $\pm$ 0.006 & 0.935 $\pm$ 0.061 & 1 \\
HD176437 & 0.753 $\pm$ 0.009 & 0.647 $\pm$ 0.117 & 4 \\
HD176437 & 0.766 $\pm$ 0.010 & 0.647 $\pm$ 0.117 & 6 \\
HD219623 & 0.542 $\pm$ 0.016 & 0.581 $\pm$ 0.013 & 3 \\
HD222368 & 1.082 $\pm$ 0.009 & 1.150 $\pm$ 0.023 & 1 \\
HD222603 & 0.581 $\pm$ 0.012 & 0.518 $\pm$ 0.014 & 3 
\enddata
\tablenotetext{}{\textbf{References.} $^{1}$\cite{Boyajian2012A}, $^{2}$\cite{Boyajian2012B}, $^{3}$\cite{Boyajian2013}, $^{4}$\cite{Maestro13},
$^{5}$\cite{Ligi12},
$^{6}$\cite{challouf},
$^{7}$\cite{Mozurkewich03},
$^{8}$\cite{Braun14},
$^{9}$\cite{Baines09},
$^{10}$\cite{Baines10}. A full version of this table is available online.}
\label{table:compare}
\end{deluxetable}

\subsection{Check on Accuracy of Derived Reddening A$_{\rm v}$}

Given that the majority ($92\%$) of our stars are within 300 pc as 
shown in Figure~\ref{fig:chi2&AVdistributions}a, we expect the median reddening of our sample to be A$_{\rm v} < 0.3$. Figure~\ref{fig:chi2&AVdistributions}b shows the reddening of our sample, with a median reddening of~A$_{\rm v, med}=\;$\medred, in good agreement with expectations. 

As a check on the accuracy of derived reddening 
A$_{\rm v}$ from \texttt{sedFit}, we use the sample of~\nstarsred~OB stars in the Cep OB3 association with directly observed UV extinction curves from the International Ultraviolet Explorer \citep[IUE,][]{Boggess78} satellite observations \citep{massa81,massa84}. We fit the SEDs of~\nstarsred~stars from photometry derived from the literature, in the same manner as \S\ref{sec:methodsdata}. Example SED fits to stars in the Cep OB3 association for HD216532 and HD216658 are shown in Figure~\ref{fig:massased}. 

We perform 160 fits for each star, with the reddening parameter spanning $A_{\rm v}=0.0-4.0$ in steps of $0.05$. For each star, similar to the methods detailed in \S\ref{sec:sedfit}, we identify the best-fit model as the one with reduced $\chi^{2}$ closest to 1.0 \citep{Andrae10}. To estimate the uncertainty on our derived reddening, we compute a $5\sigma$ confidence interval, identifying models where $\Delta\chi_{\rm 99.73\%}^{2} = 11.8 < \chi^{2}-\chi^{2}_{\rm best}$, where $\chi^{2}_{\rm best}$ corresponds to the the best-fit model, and $\Delta\chi_{\rm 99.73\%}^{2}=11.8$ corresponds to models that fall within the $5\sigma$ confidence interval \citep[see section 15.6,][]{Press02}. 

As shown in Table~\ref{table:massa} and Figure~\ref{fig:massa}, we find our fitted A$_{\rm v}$ for \nstarsred\ stars in good agreement with the directly observed~A$_{\rm v}$ for each star from \cite{massa84} (c.f.\ 4th column, Table 2), with a mean offset in the reddening of \massaoff~mag,demonstrating good agreement. This small systematic offset in reddening should not significantly affect the accuracy of the derived angular diameters of our calibrator stars, as we now show.

\section{Discussion: Comparison to Previous Catalogs}
\label{sec:disc}

Several authors have previously presented catalogs of interferometric calibrators, including e.g., 
\citet{Borde:2002}, \citet{Merand:2005}, \citet{Bonneau:2006}, \citet{vanBelle:2008}, and \citet{Richichi09}. These catalogs were optimized for infrared interferometry, as appropriate for infrared beam combiners on the Very Large Telescope Interferometer and the CHARA array. Our catalog builds on these previous works, and extends them by optimizing a calibrator sample for the optical, as is needed by next-generation interferometers such as VISION at NPOI \citep[e.g.,][]{Garcia16}. Furthermore, of the~\totstars~ stars in our sample, there are only 97 stars in common with the \citet{Merand:2005} catalog, while only 79 overlap with \citet{Borde:2002}, showing we have greatly expanded the potential calibrator pool. In Figure~\ref{fig:disc} we compare our angular diameter estimates to those of \citet{Merand:2005} and \citet{Borde:2002}, finding them in good agreement with no statistically significant systematic offset. These works focused on larger potential calibrator stars that were optimal for 200-m class (and smaller) interferometers working in IR bands, while the vast majority of our catalog contains stars smaller than 1.0 milliarcsecond (Figure~\ref{fig:Diamdistributions}), which are ideal for the longest ($>300$-m) baselines on next-generation optical interferometers.

For our purposes here, we can also utilize previous interferometric observations as a check on the accuracy of our limb-darkened angular diameter estimates, $\theta_{\rm LD}$. Specifically, we compare \starscompared\ stars from our list to their interferometrically measured diameters from the literature \citep{Mozurkewich03,Baines09,Baines10,Ligi12,Boyajian2012A,Boyajian2012B,Boyajian2013,Maestro13,challouf,Braun14} in Table~\ref{table:compare} and Figure \ref{fig:SEDvsMEAS}. Our angular diameter estimates are in excellent agreement with those measured interferometrically, with a mean offset of~\avgoffset~$\pm$~\stddevoffset~mas. There is no systematic offset between our estimated calibrator star angular diameters from \texttt{sedFit} and their corresponding measured angular diameters from the literature.

\section{Summary}
\label{sec:conclusion}
Of the $\sim\;$3000 stars in the northern hemisphere brighter than V$<6$ and with declinations $-15^{\circ}<\delta<82^{\circ}$, we have compiled a list of \totstars\ stars well suited as interferometric calibrators, with \textit{uniformly} estimated angular diameters and bolometric fluxes from SED fitting. Each star was fit with a model SED, including a reddening correction, to literature photometry spanning $0.2-10$~$\mu$m, using a spectral template on input, corresponding to a specific spectral type. We purged this list of known binary and multiple stellar systems, stars with poor-quality or sparse photometry, and stars where the spectral template fit resulted in a poor reduced $\chi^2$.

For a subset of the stars in our sample, our derived angular diameters show excellent agreement with directly measured angular diameters from the literature, and with diameter estimates appearing in previous calibrator catalogs. In addition, our derived reddening for OB stars in the Cep OB3 association show excellent agreement with directly measured reddening for those stars from the literature.

With the next generation of interferometers that operate at visible wavelengths now operating and more coming online, this catalog should serve as a broadly useful resource for years to come.

\section*{Acknowledgements}
We thank the anonymous referee for helpful critiques that improved the clarity and quality of this paper. S.J.S.\ would like to thank the NSF REU program at Vanderbilt University for the opportunities and support that made this work possible.
E.V.G.\ and K.G.S.\ gratefully acknowledge partial support from NSF PAARE grant AST-1358862.
This research has made use of the SIMBAD database,
operated at CDS, Strasbourg, France. 
This research has made use of the Washington Double Star Catalog maintained at the U.S.\ Naval Observatory.

\bibliography{report}

\end{document}